\documentclass{article}
\usepackage{spconf,amsmath,graphicx}

\title{Time-Sequence Channel Inference for Beam Alignment\\ in Vehicular Networks}
\name{Sheng Chen, Zhiyuan Jiang, Sheng Zhou, Zhisheng Niu, \emph{Fellow, IEEE}\thanks{This work is sponsored in part by the Nature Science Foundation of China (No. 61701275, No. 91638204, No. 61571265, No. 61621091), the China Postdoctoral Science Foundation, and Intel Collaborative Research Institute for Mobile Networking and Computing. Emails: \{chen-s16@mails., zhiyuan@, sheng.zhou@, niuzhs@\}tsinghua.edu.cn.}}
\address{Beijing National Research Center for Information Science and Technology,\\ Tsinghua University, Beijing 100084, China. }

\begin{document}
\maketitle

%%%%%%%%%%%%%%%%%%%%%%%%%%%%%%%%%%%%%%%%%%%%%%%%%%%%%%%%%%%%%%%%%%%%%%%%%%%%%%%%
\begin{abstract}

In this paper, we propose a learning-based low-overhead beam alignment method for vehicle-to-infrastructure communication in vehicular networks. The main idea is to remotely infer the optimal beam directions at a target base station in future time slots, based on the CSI of a source base station in previous time slots. The proposed scheme can reduce channel acquisition and beam training overhead by replacing pilot-aided beam training with online inference from a sequence-to-sequence neural network. Simulation results based on ray-tracing channel data show that our proposed scheme achieves a $8.86\%$ improvement over location-based beamforming schemes with a positioning error of $1$m, and is within a $4.93\%$ performance loss compared with the genie-aided optimal beamformer. 
\end{abstract}


%%%%%%%%%%%%%%%%%%%%%%%%%%%%%%%%%%%%%%%%%%%%%%%%%%%%%%%%%%%%%%%%%%%%%%%%%%%%%%%%
\section{INTRODUCTION}
~~~~~Recently, vehicular networks have drawn great attention from both industry and academia for its significant potential in many real-world applications \cite{gerla2011vehicular}. In vehicular networks, vehicles can communicate with, among others, roadside units (RSUs), which is referred to as vehicle-to-infrastructure (V2I) communications \cite{gozalvez2012ieee}. In general, V2I communications require high-rate data links due to the massive data requirements from vehicle applications, e.g. the acquisition of high precision maps. Besides, timely message dissemination for security issues brings about the challenge for ultra-reliable and low-latency communications.

Multiple-input multiple-output (MIMO) is considered as an indispensable technology for future communication systems \cite{shafi20175g}. To achieve the benefits of spatial diversity and multiplexing gains in MIMO systems, timely and accurate channel state information (CSI) is needed at both sides of the transceivers. In addition, the usage of large antenna arrays makes it possible to utilize higher frequency bands, e.g. millimeter wave bands \cite{roh2014millimeter}. In mm-wave enabled vehicular networks, the RSUs need to adjust the beam directions towards the vehicles before data transmissions in order to achieve beamforming gain which is essential for mm-wave transmissions. Hence efficient beam training and tracking methods are needed, i.e., beam alignment \cite{song2013millimeter}. Existing exhaustive search \cite{payami2015effective} and hierarchical search \cite{alkhateeb2014channel} schemes suffer from high training overhead, degrading the quality of service for V2I communications. furthermore, the high mobility of vehicles leads to frequent handovers between RSUs, leading to more frequent beam alignment in vehicular networks. As a result, novel beam alignment methods with low overhead are in urgent need.

\begin{figure}[!t]
	\centering
		\includegraphics[width=0.45\textwidth]{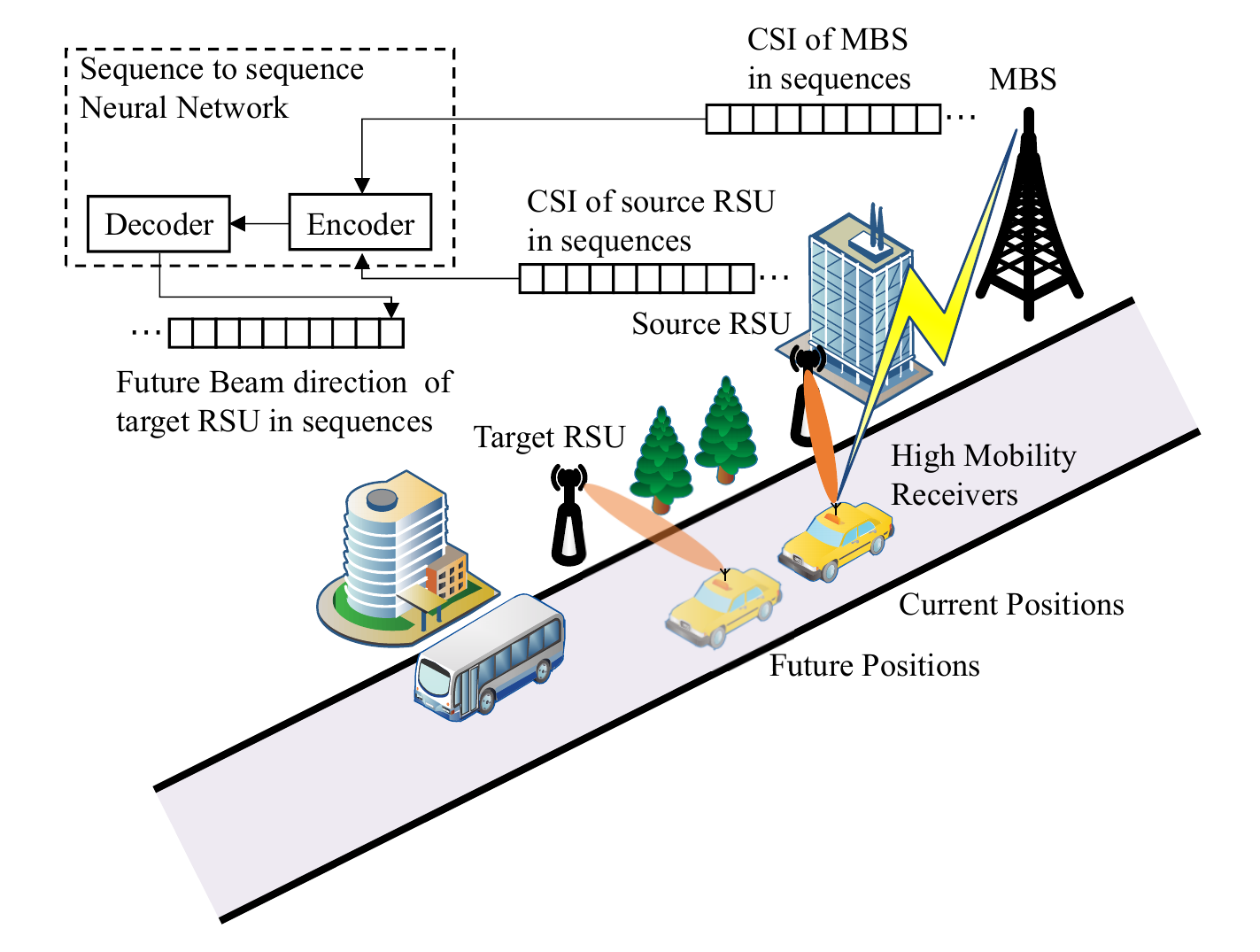}
	    \caption{Schematic of the proposed channel inference based beam alignment method.}
	    \label{fig_sce}
\end{figure}

Extensive work has been devoted for fast and efficient beam alignment. Some work considers using location information to help infer the beam patterns \cite{kela2016location}. However, the acquisition of location information brings extra overhead and the performance is highly related to the positioning accuracy. Authors in \cite{li2017super,zhu2018high} proposed beam tracking methods, but these methods can hardly deal with the handover case, for the reason that the channel state of the target BS cannot be easily estimated based on the observation of the source BS channels. In our previous work \cite{liu2015seeing,chen2017remote}, the non-linear correlation between the channels of neighbouring BSs was studied and a remote beamforming inference scheme was proposed; specifically, the beam directions of small BSs can be learned from the CSI of the macro BS (MBS). Inspired by that, we propose a learning-based beam alignment method for V2I communication in vehicular networks. The basic idea is to infer the beam direction of the target BS, from the CSI of the source BS. Here the target BS is the RSU to be switched to, while the source BS can be the RSU that serves the vehicle currently,  or the covering MBS in a heterogeneous scenario. The main challenge is that the estimation delay during the prediction process has a significant influence on the beamforming performance when the receivers are of high mobility. Hence, we adopt a sequence to sequence channel learning framework to predict the beam directions in advance, where the learning process is done by a Deep Neural Network (DNN) with encoder-decoder architecture.  

The main contributions of this work are as follows: 

1) We design a novel beam alignment method with low signaling and time overhead for V2I communications in vehicular networks.

2) We propose a time-sequence channel learning framework, which is able to continuously predict the future beam directions of the target BS, based on the CSI of the source BS in previous time slots. 

3) The performance of the proposed scheme is evaluated on ray-tracing channel data, and the simulation results of the proposed approach outperform the location-based beamforming even without positioning error.

\section{System Model}
~~~~~As shown in Fig. \ref{fig_sce}, a vehicle will be switched to the target RSU from the source RSU. In order to save the beam training overhead for the target RSU, our proposed framework can infer the beam directions of the vehicle based on the known CSI from the source RSU or the MBS. The high-level procedures are as follows: when handover occurs,  the source RSU predicts the beam directions of the target RSU in future time slots based on the CSI of its own in previous time slots, and sends the predicted beam directions to the target RSU through the backhaul link. The target RSU can first measure the inference and transmission delay by using  timestamps, and then establish the link to the vehicles using the estimated beam directions with the measured delay. If the MBS can continuously collect the CSI of the vehicle, the CSI of MBS can also be utilized to help infer the beam directions of the target RSU, and the procedures are similar to the aforementioned method learning from source RSU.
		
The main challenge is to find the mapping function from the CSI of the source BS in previous time slots to the beam directions of the target BS in the following time slots. We resort to a data-driven approach, which is to build a learning model to approximate the mapping function in a supervised way. In other words, the target is to find appropriate function $f$ and parameters $\theta$ holding the following formula:
\begin{equation}
\label{eqn1}
    f([\boldsymbol{h}_{-T},\cdots,\boldsymbol{h}_{-1},\boldsymbol{h}_{0}],\boldsymbol{\theta}) = [\boldsymbol{\hat{d}}_1,\boldsymbol{\hat{d}}_2,\cdots,\boldsymbol{\hat{d}}_K],
\end{equation}
where $\boldsymbol{h}_{-T},\cdots,\boldsymbol{h}_{-1}$ denote the CSI of the source BS in past $T$ time slots, $\boldsymbol{\hat{d}}_1,\boldsymbol{\hat{d}}_2,\cdots,\boldsymbol{\hat{d}}_K$ denote the selection probability of each beam direction for the target BS in future $K$ time slots. The optimization target is the cross-entropy between the estimated selection probability $\boldsymbol{\hat{d}}$ and the one-hot encoded vector of the optimal beam direction $\boldsymbol{d}$, which can be expressed as:

\begin{equation}
    \min -\frac{1}{K}\sum_{k=1}^{K} \sum_{x=1}^{X} \boldsymbol{d}_k(x)\log\boldsymbol{\hat{d}}_k(x),
\end{equation}
where $x$ indicates the beam directions.

Assuming that both the MBS and the RSUs are equipped with uniform linear antenna array, the steering vector of those BSs can be expressed as:
\begin{equation}
 [\textbf{a}(\theta)]_i = e^{-j\frac{2\pi}{\lambda} id\sin \theta} (i = 0,1,\cdots,N-1),
\end{equation}
where $\lambda$ denotes the wavelength, $d$ denotes the antenna spacing, $N$ denotes the number of antennas, $\theta$ denotes the angle of departure (AoD) for transmitters or the angle of arrival (AoA) for receivers.

Considering a narrow band scenario, the downlink CSI between the BSs and the UEs can be expressed as:
\begin{equation}
\label{eqn4}
\textbf{H} = \sum_{i=1}^{N_\textsc{p}} \alpha_i \textbf{a}_\textsc{B}(\theta_{di})\textbf{a}_\textsc{U}^T(\theta_{ai}),
\end{equation}
where $\alpha_i$, $\theta_{di}$, $\theta_{ai}$ denotes the complex impulse response, DoD, DoA of the $i$-th propagation path between transceivers, respectively,  $N_\textsc{p}$ denotes the total number of propagation paths.

\section{Time-sequence Channel Learning Framework}
~~~~~In this section, the architecture of the proposed sequential channel learning framework will be introduced, and some practical analysis will be given.

\subsection{Network Architecture}
~~~~~Neural Network (NN) has shown powerful strength to approximate nonlinear functions and made great achievements in many machine learning problems \cite{hornik1989multilayer}. Recurrent Neural Network (RNN) is an important NN architecture for sequential data analysis, which is also widely used in Natural Language Processing problems \cite{lipton2015critical}. Notice the fact that both the input and the output of our problem are sequential channel data. Hence it's a natural idea to build a sequence to sequence framework \cite{sutskever2014sequence}. Specifically, the framework consists of an encoder and a decoder, where the encoder maps the input sequences to a vector and the decoder translates the vector to the target sequences. 

%\begin{figure*}[!t]
%	\centering
%		\includegraphics[width=0.9\textwidth]{NN_arch.pdf}
%	    \caption{The architecture of the proposed sequential channel learning framework.}
%	    \label{fig_rnn}
%\end{figure*}

\begin{figure}[!t]
	\centering
		\includegraphics[width=0.45\textwidth]{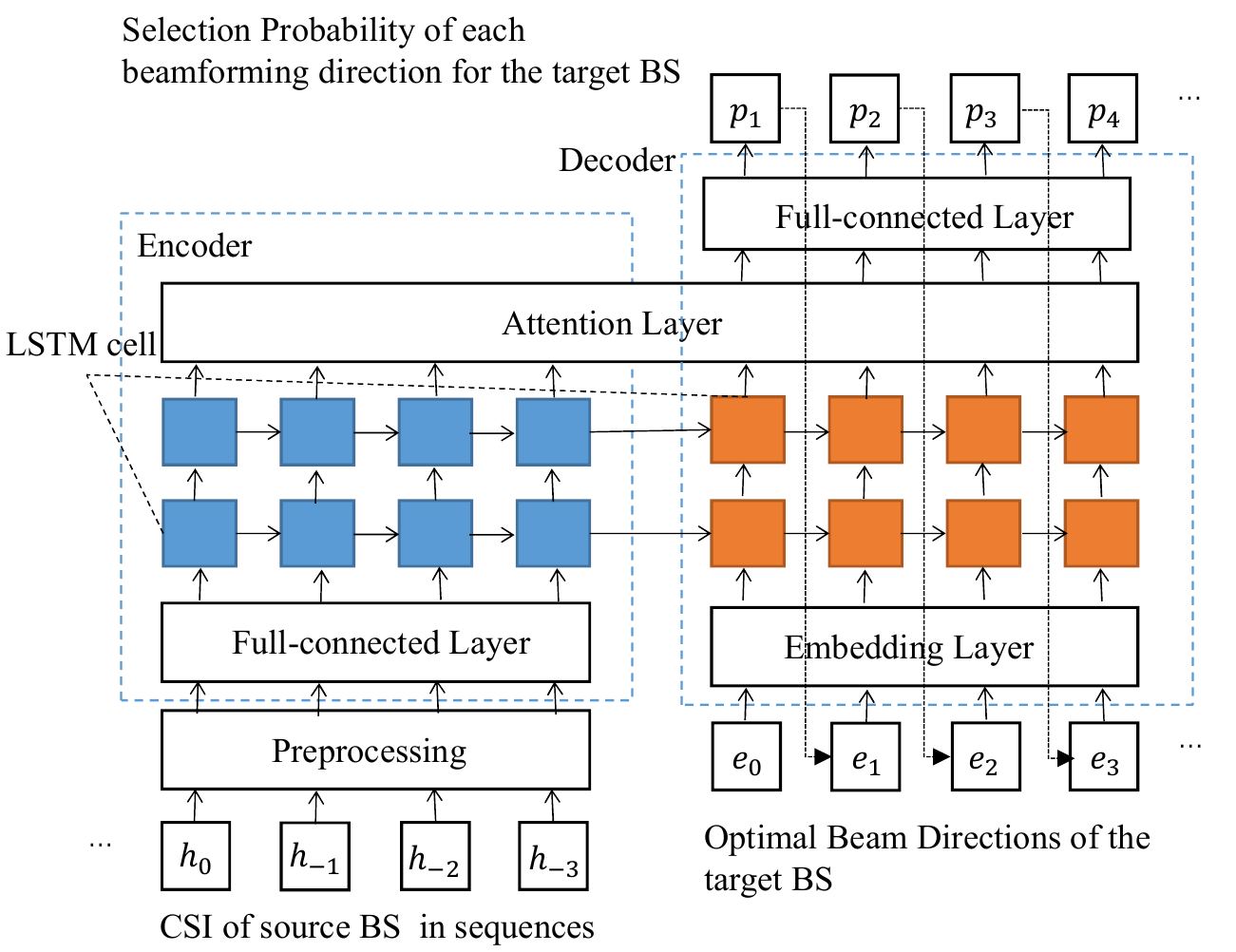}
	    \caption{The architecture of the proposed time-sequence channel learning framework.}
	    \label{fig_rnn}
\end{figure}

Fig.\ref{fig_rnn} shows the architecture of the proposed learning framework. The inputs of the encoder are the pre-processed channel sequences of the source BS. In order to help accelerate the training speed of the NN, we first change the CSI into angular domain by using fast Fourier transformation, then extract the amplitude information and finally take the logarithm. The encoder consists of a full-connected layer with $256$ hidden nodes, two long short term memory (LSTM) layers with $256$ hidden nodes each layer \cite{hochreiter1997long}, and one attention layer. Here we adapt the attention model proposed in \cite{luong2015effective} to further improve the estimation accuracy. 

As for the decoder, the outputs are the selection probabilities for different beam directions of the target BSs in future time slots, while the inputs are different for the training and inference process. During the training process, the inputs are the target indicators of the optimal beam directions in future time slots $e_1,e_2,\ldots$, adding a start token $e_0$ at the head of the sequences. During the inference process, the input of the decoder starts from the start token $e_0$, and the input in next slot is based on the output of the decoder in current time slot. The embedding layer maps the input indicator to a $1\times 100$ vector, following by two LSTM layers with $256$ hidden nodes as the hidden layers. The hidden value of the decoder is initialed from the last hidden state of the encoder. 

Dropout is utilized here in order to prevent the trained model from over-fitting \cite{srivastava2014dropout}. The cost function is the cross-entropy between the estimated selection probability of future beam directions and  the one-hot encoded vectors of the optimal beam directions. Adam optimizer is applied for updating the gradients of the parameters in the NN \cite{kingma2014adam}.

\subsection{Practicality Analysis}
~~~~~The training data can be collected with the help of the conventional beam alignment methods. For example, the target BS can measure the optimal beam directions of users by beam search and feed them back towards the source BS. Combined with the estimated CSI of the source BS in previous time slots, training data can be collected and utilized for updating the learning model. The training process can be done offline, hence the online inference cost is to feed one input sequence into a NN and obtain the output sequence. The calculation overhead is acceptable, while the time cost can be compensated by prediction in advance.

\section{Simulation Results}
\begin{figure}[!t]
	\centering
		\includegraphics[width=0.45\textwidth]{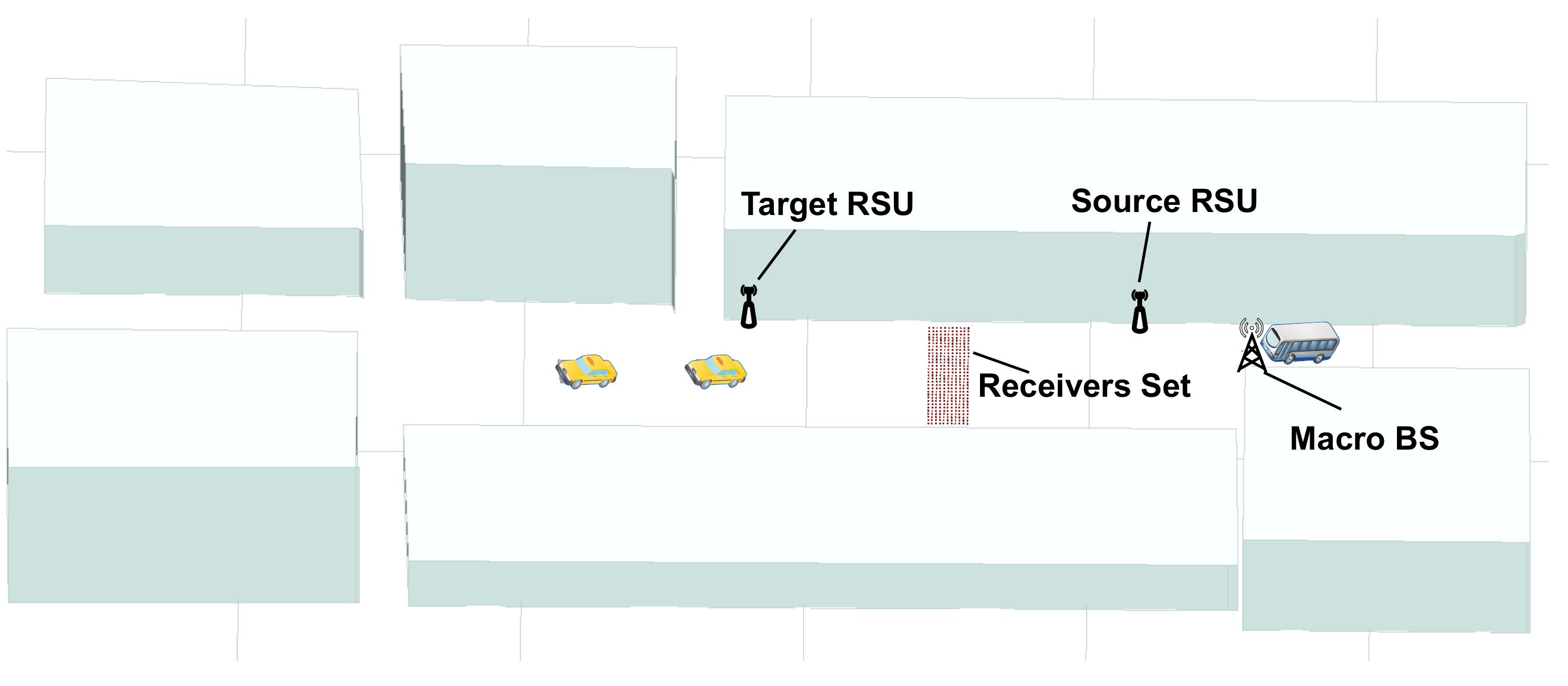}		
	    \caption{Simulation scenarios.}
	    \label{fig_sim_set}
\end{figure}

~~~~~The performance of the proposed scheme is evaluated on ray-tracing channel data generated by wireless incite. As shown in Fig. \ref{fig_sim_set}, two RSUs are within the coverage of one MBS. The RSUs are equipped with a $32\times1$ linear antenna array, while the MBS is equipped with a $128\times1$ linear array. The heights for the RSUs and the MBS are $3$~meters and $22$~meters, respectively. The receivers are equipped with single antenna, located at a $10\times30$~meters uniform grid in the edge of two RSUs. The spacing of two neighbouring points in the grid is $0.05$~m, and the CSI of the vehicles can be approximated by the CSI of the closest sampling point in the receivers set. The CSI of the sampling points can be calculated by firstly measuring the complex impulse response, the AoA and the AoD of each ray, and then substituting to (\ref{eqn4}). The central point of the frequency band is $28$~GHz. The vehicles are assumed to be running on a straight line with fixed acceleration, and the time interval is set to be $1$~ms. The initial speed for each vehicle is uniformly distributed between $10$~ms$^{-1}$ and  $15$~ms$^{-1}$, while the acceleration is uniformly distributed between $-3$~ms$^{-2}$ and  $3$~ms$^{-2}$.

We adopt a $256$-dimension DFT matrix as the beamforming codebook, which can be easily realized by a analog beamformer. Since the RSUs only have $32$ antennas, we extract the first $32$ dimensions of the DFT matrix as the codewords. In the simulation settings, we feed the channel sequences of the source BS in the past $50$ ms to the NN, and the optimal beam directions of the target BS in the following $50$ ms can be inferred. We compare the results of our scheme with two baseline algorithms. The first baseline is location-based beamforming, where the beam directions of the vehicles are estimated by calculating the AoD of the direct link based on the location of transceivers. The second baseline is the method proposed in \cite{chen2017remote}, where the beam directions of the target RSUs can be learned by the out-dated CSI of the source RSUs using a feed-forward NN with 4 hidden layers.

\begin{figure}[!t]
	\centering
		\includegraphics[width=0.45\textwidth]{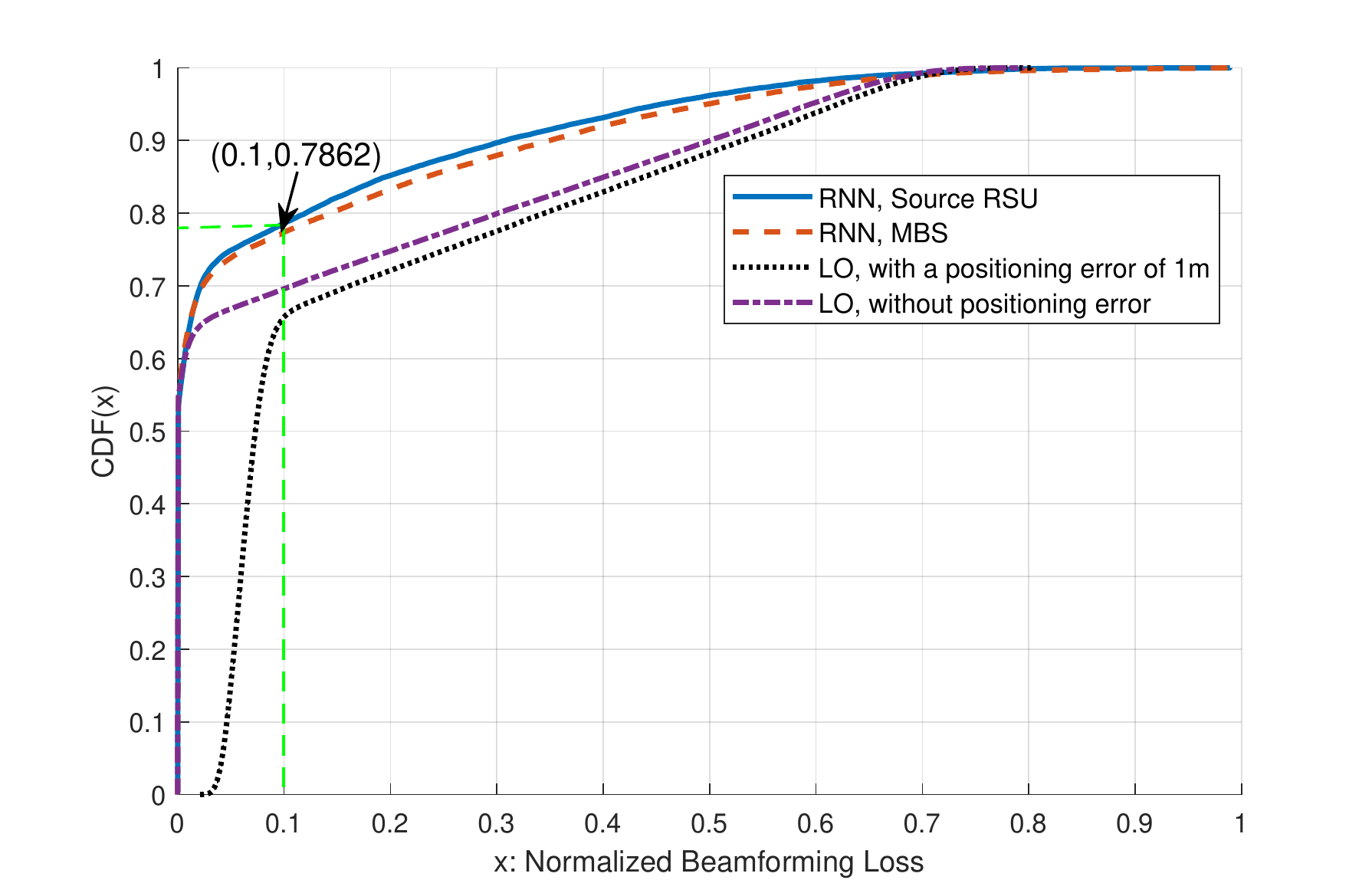}		
	    \caption{CDF of the normalized beamforming loss; RNN stands for the proposed NN based approach, while RSU and MBS denote learning from the CSI of source RSU and MBS, respectively ; LO stands for location based beamforming; }
	    \label{fig_cdf}
\end{figure}

The cumulative distribution functions (CDF) of the normalized beamforming loss for different schemes are shown in Fig. \ref{fig_cdf}. Here the normalized beamforming loss is defined as the normalized distance between the received signal strength (RSS) using predicted beam directions and the RSS using the optimal beam codeword from the predefined codebook. It can be seen that the probability of the beamforming loss smaller than $10^{-1}$ is $78.62\%$ by learning from the Source RSU. Addtionally, simulation results show that learning from either the neighbouring RSU or the MBS can provide a better performance compared with location based beamforming with even no positioning error.

\begin{figure}[!t]
	\centering
		\includegraphics[width=0.45\textwidth]{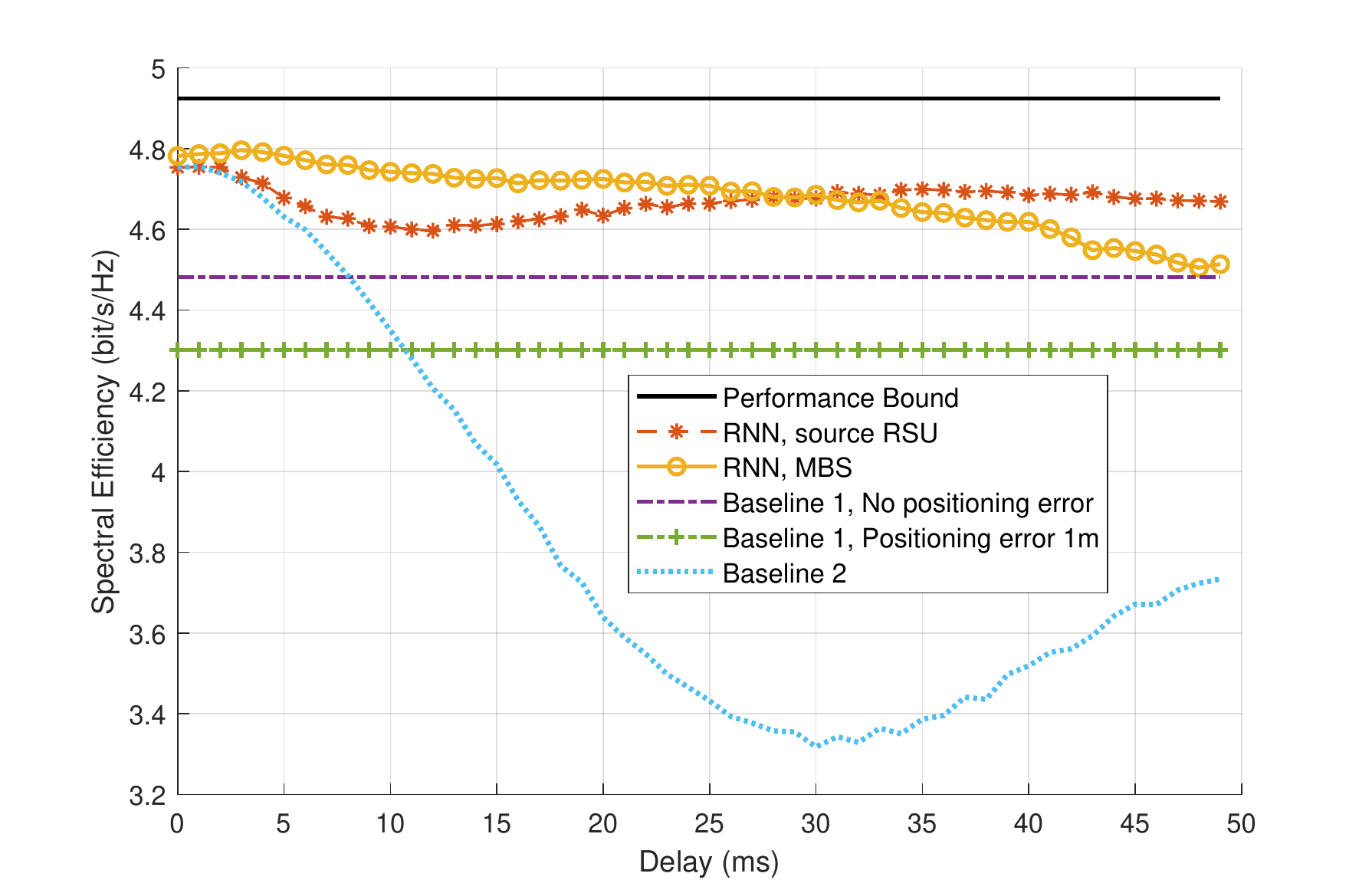}		
	    \caption{Spectral efficiency versus prediction delay; RNN stands for the proposed scheme, while RSU and MBS stands for different source BSs.}
	    \label{fig_spec}
\end{figure}

Fig. \ref{fig_spec} shows the spectral efficiency of different schemes considering the prediction delay. It can be observed that our proposed algorithms have a $8.86\%$ improvement over the location based algorithm with a positioning error of $1m$, and a $4.93\%$ performance loss compared with the performance bound, which is defined as the spectral efficiency using the optimal beamforming vector from the codebook. Besides, the second baseline  can achieve an equivalent performance over our proposed scheme when the prediction delay is small. However, the performance drops significantly when the delay is getting larger, while our proposed method can predict the future beam directions very well.

\section{CONCLUSIONS}
~~~~~In this paper, we propose a time-sequence channel learning framework for beam alignment in vehicular networks. A sequence to sequence neural network is built to predict the future beam directions of the target BS based on the previous CSI of the source BS. Performance of the proposed scheme is evaluated on ray tracing based channel data. Simulation results show that the proposed scheme is within a $4.93\%$ performance loss compared with the genie-aided optimal beamformer. 

An interesting future direction is to consider multi-user beamforming. Currently only the optimal beam direction of a single user can be learned, neglecting any multi-user interference. To achieve multi-user spatial multiplexing gain, more information about the channel state needs to be learned. On the other hand, an online training approach is beneficial if time-varying scattering environments are considered.


%%%%%%%%%%%%%%%%%%%%%%%%%%%%%%%%%%%%%%%%%%%%%%%%%%%%%%%%%%%%%%%%%%%%%%%%%%%%%%%%

%%%%%%%%%%%%%%%%%%%%%%%%%%%%%%%%%%%%%%%%%%%%%%%%%%%%%%%%%%%%%%%%%%%%%%%%%%%%%%%%

%%%%%%%%%%%%%%%%%%%%%%%%%%%%%%%%%%%%%%%%%%%%%%%%%%%%%%%%%%%%%%%%%%%%%%%%%%%%%%%%
%\section*{ACKNOWLEDGMENT}

%%%%%%%%%%%%%%%%%%%%%%%%%%%%%%%%%%%%%%%%%%%%%%%%%%%%%%%%%%%%%%%%%%%%%%%%%%%%%%%%
\bibliographystyle{ieeetr}
\bibliography{seqbeam}
\end{document}